# Character comes from practice: longitudinal practice-based ethics training in data science

Louise Bezuidenhout[1]
Emanuele Ratti[2]

**Abstract**: In this chapter, we propose a non-traditional RCR training in data science that is grounded into a virtue theory framework. First, we delineate the approach in more theoretical detail, by discussing how the goal of RCR training is to foster the cultivation of certain moral abilities. We specify the nature of these 'abilities': while the ideal is the cultivation of virtues, the limited space allowed by RCR modules can only facilitate the cultivation of superficial abilities or proto-virtues, which help students to familiarize with moral and political issues in the data science environment. Third, we operationalize our approach by stressing that (proto-)virtue acquisition (like skill acquisition) occurs through the technical and social tasks of daily data science activities, where these repetitive tasks provide the opportunities to develop (proto-)virtue capacity and to support the development of ethically robust data systems. Finally, we discuss a concrete example of how this approach has been implemented. In particular, we describe how this method is applied to teach data ethics to students participating in the CODATA-RDA Data Science Summer Schools.

**Keywords**: Responsible Conduct of Research; Virtue Ethics; Micro-ethics; Data Science; Open Science

1. INTRODUCTION

In the past few years, data science has spread across virtually all working, managerial, and research environments. The tools provided by this new interdisciplinary endeavor have made possible the analysis of large data sets, allowing the actionable potential of Big Data to be unleashed. Despite the advantages already realized by the data revolution, advances have been accompanied by the emergence of complex ethical, political, and societal issues. Unsurprisingly, scientific research using data science tools is not immune from these issues. For this reason, there is increasing pressure on academic institutions to promote the responsible use of data science tools in research.

---

[1] Center for Science and Technology Studies (CWTS), Leiden University
[2] Department of Philosophy, University of Bristol

Many institutions formulating training in data ethics are doing so in the context of Responsible Conduct of Research (RCR) programs. In these cases, RCR content is adapted to the specificities of data science by the inclusion of discipline-specific case studies. While inheriting the strengths of the RCR approach, these programs also inherit much of the flaws that have led to criticism. These include the promotion of a 'compliance-based' approach to good conduct, and an emphasis on ethics as a bureaucratic activity.

In this article, we propose an ethics program for data scientists framed in contrast to traditional RCR programs. We outline a new framework for a non-traditional data science RCR based on the language of virtue ethics and the notion of microethics. This new framework is capable of supporting the aspirational nature that motivated the creation of RCR programs decades ago, and fostering contextual awareness and strategies of resilience amongst data scientists.

The structure of this article is as follows. In Section 2, we describe the nature of traditional RCR programs, and their flaws (2.1). In addition, we show why data science results in new issues that require RCR training (2.2), and why we cannot use traditional RCR programs as blueprints (2.3). In Section 3, we describe a framework for a new RCR program tailored for data science. Finally, in Sections 4 and 5 we describe how our RCR framework works, by describing its application in five steps, exemplified by the Committee on Data Research and Research Data Alliance (CODATA-RDA) Schools for Research Data Science.

## 2. RCR TRAINING AND DATA SCIENCE

### 2.1 RCR Programs and Their Flaws

RCR is a widespread institutional training that aims to familiarize scientists with ethical, legal, and societal issues arising within scientific research. In the United States context, RCR training was mandated by the NIH in 1989, and by the NSF in 1997 (Chen 2021). RCR training utilizes a framework that outlines areas in which researchers have responsibility. These include mentoring, authorship, conflicts of interests, protection of human subjects, use of animals in research, as well as "FFP" (fabrication, falsification and plagiarism) misconduct (Shamoo and Resnik 2015). While the RCR discourse has been spelled out in various ways, especially in the research context of the United States principlism has been seen as a mean of structuring ethical discussion (Shamoo and Resnik 2015). Despite the varieties of ways in which these programs are designed, the philosophy and the roll-out of RCR has been associated with a number of tensions. Chen (2021) elaborates on these issues, which we summarize as follows.

The first issue is that mandated RCR training has not been accompanied by mandated funding allocations, meaning that institutions dedicate varying amounts of resources to the development and roll-out of training. This has led to considerable variation in the way the content is delivered. In most cases, well-established RCR programs reflect institutional champions and supportive leadership, rather than simply the size of institutions and institutional budgets.

The variations in RCR roll-out can be seen as a side-effect of a lack of clarity on what RCR should accomplish. As Chen (2021) shows, an assessment of current RCR training shows a high degree of variability in terms of expected outcomes - from rule compliance to character change. Similarly, methods for evaluating RCR training outcomes are highly variable, making cross-institutional comparisons challenging. Moreover, responsibility for RCR training, position of RCR training on institutional budgets, and accountability for long-term assessment differ across institutions.

The roll-out of RCR training, challenging as it is, masks a number of tensions and paradoxes that persist within RCR as a subject. Many scholars recognize the mandatory nature of RCR training as a primary source of problems. On the one hand, RCR programs were motivated originally by appealing to an aspirational component, "seeking to promote research integrity through changing behavior, character, and judgment" (Chen 2021, p. 232). This is consistent with an 'authenticity paradigm of professional ethics' (Kelly 2018): the efforts to initiate and develop RCR training exemplify on-going 'existential' discourses of scientific communities, aiming at self-understanding, most notably in the form of delineating their own 'ethos', goals, and relations with the public. On the other hand, the mandatory nature of RCR programs has promoted a 'compliance paradigm' (Kelly 2018) that conflates with a 'rule-following' dimension. This can lead researchers to interpret RCR as "meaningless boxes to be checked, (…) byzantine and vaguely threatening bureaucracies" (Chen 2021, p. 231). This bureaucratic interpretation of RCR training misrepresents ethics as external to scientific practice, and not as something that can be learnt in an analogous way in which scientific skills are learnt. Such training represents an ethical researcher as following well-defined rules, thus conflating unquestioning compliance and ethics. This approach obviously contrasts to the original aspiration of promoting "deep and thoughtful exploration into the various contexts and factors at play in ethical issues" (Chen 2021, p. 231).

The challenges embedded within current RCR framings and roll-out has lessened the potential training to function as a transformative tool to build character. This lost opportunity, while often remarked on, is difficult to address within the current structures of RCR training.

**2.2 Why We Need RCR Trainings for Data Science**

Data science has recently emerged as an interdisciplinary endeavor. This rapidly evolving discipline integrates practitioners from different scientific and engineering traditions (e.g., statistics, computer science, mathematics, etc.) to generate, process and visualize research data. As a result, researchers occupying data science roles are heterogeneous in terms of disciplinary expertise and computing training. Because of this, it is difficult to understand the boundaries between data science and the other fields that are 'assembled'. Who, exactly, is a data scientist remains an open question.

Despite its heterogeneity in application, data science coalesces around a number of common practices. The use of computer programming to conduct data analyses means that – regardless of the application – data scientists often make use of a common set of computer programs and programming languages. These common practices also give rise to a common 'way of doing', in that daily data science activities are comprised of a number of repeated tasks such as cleaning, annotating and visualizing data, as well as writing code to assist these activities. Data science, it can be said, is both discrete and iterative.

There has been plenty of discussions about data science not just for its interdisciplinary and 'vague' (in the technical philosophical sense) nature. In fact, it has been overwhelmingly documented that data science raises plenty of moral and political issues relating to data creation, analysis and re-use. We understand 'moral' as how the design and use of data science tools influence data subjects' lives and their autonomous choices (Ratti and Graves 2021), while 'political' refers to how the design and use of data science tools end up amplifying those 'moral issues' at a societal/structural level. For instance, it has been shown that large language models (LLM) – used in heterogeneous contexts from designing chatbot to the early phases of drug

design – can incorporate intrinsic biases in the forms of misrepresentation, underrepresentation, and overrepresentation (Bommasani et al 2021). These intrinsic biases can potentially lead to extrinsic harms, which can be at the individual level, and especially at the collective level, when for instance they result in representational harms effecting both performance disparities among groups, as well as a technical standardization of biases (Bommasani et al 2021). These problems have put a lot of emphasis on underappreciated, highly technical and repetitive practices such as data curation and documentation, or model tuning (Bender et al 2021). These issues are particularly pressing, and they also apply to scientific research, in particular in fields such as climate science or biomedicine, where the use of data science tools is spreading rapidly. The spreading of data science can be direct or sometimes indirect: it can be by using some of its tools, or simply by taking advantage of the infrastructures (e.g., databases) that this new field has made possible.

Key to understanding the ethical, political and social implications of data science tools are two fundamental issues. The first is the realization that all algorithmic systems, as products of human action, are *value-laden*. This links to the problem of algorithmic bias (Fazelpour and Danks 2021). 'Bias' here "simply refers to deviation from a standard (…) [where] there are many types of bias depending on the type of standard being used" (Danks and London 2017, p. 4692). Accordingly, an algorithm can be morally, statistically, or socially biased. Algorithms are optimized "for performance relative to a standard" (Fazelpour and Danks 2021, p. 3), where such standards are built upon preferences which, at their core, are values. To illustrate this point with a straightforward example, consider this. When data scientists are cleaning data sets that are used to train an algorithm, they often deal with OCR data, which are very time-consuming to prepare because they are full of gaps. For this reason, data scientists may prefer data that have been gathered digitally. Behind this choice, there is a preference for speeding up routine operations like data cleaning and preparation motivated by efficiency, as well as a preference for high quality data, which can be interpreted as a 'desideratum' or a value for data. However, in the health care context systematically using only high-quality data may result in using data only from wealthy areas, where healthcare infrastructures have the capability of gathering data in a more cutting-edge way. This will result in neglecting data from underserved areas, where data gathering is done by using more traditional means because of the lack of resources (Ratti and Graves 2021). This, long term, may result in representational harms, especially in segregated countries like the United States. It is important to recognize that these biases are not necessarily the byproduct of malicious behavior. In fact, it is difficult to point to cases where the biases are intentional. They can happen independently of the intentions of practitioners when, for instance, a particular framing is used that systematically biases the entire system towards a certain value-laden direction (Selbst et al 2019). In other words, algorithms are biased in the sense that they are value-laden, and as such they can be socially and morally controversial. In addition to the example just mentioned, one can also consider other famous examples such as the algorithmic tools to calculate the probability of recidivism (ProPublica 2016), which seem to incorporate specific notions of what is the goal of justice in the justice system (Pruss 2020). While algorithmic biases have been widely discussed in relation to society-focused computing, such as facial recognition software, there are additional potential biases within academic research. As a result, each data scientist has the potential to integrate algorithmic biases into analysis and visualization software.

The second way in which data science tools can be morally or socially problematic is that they generate ethical 'effects' when they are deployed. For instance, when algorithmic systems are used and scaled-up, they constrain decisions and capabilities (Ratti and Graves 2021). This can lead to unexpected downstream ethical consequences that may not match practitioners' intentions. In the case of cleaning data mentioned above, focusing only on high quality data may result in promoting overrepresentation of certain social groups, while promoting underrepresentation of others. It is unlikely that a data scientist who is more comfortable in dealing with high quality data will promote such dynamics of under- and overrepresentation on purpose. In other words, there can be ethical effects that are very difficult to anticipate, and to connect to routine and technical practices involving data. We call these 'unanticipated effects'. Therefore, data scientists have been wondering how exactly to scale such tools in a way that minimizes such ethical effects (Bommasani et al 2021).

The presence of values in algorithms and unanticipated effects pervade the daily activities of data scientists. It has been convincingly documented how micro- and incremental technical choices are inevitably intertwined with values, and can have ripple-effects, such as the way health-risks are managed at the population level (Obermeyer et al 2019). This can happen in virtually any phase of the data science process, including the definition of a problem to be solved by the algorithm (Fazelpour and Danks 2021; Selbst and Barocas 2018), data acquisition and preparation (Ratti and Graves 2021), data analysis and modeling (Graves and Ratti 2021), and deployment (Wachter, Middelstadt and Russel 2021).

What is needed, then, is a means of teaching ethical awareness and sensitivity to data scientists that foregrounds the cumulative nature of data science. This means that data scientists need to recognize that ethically sound digital infrastructures and data analyses are influenced by the micro-technical decisions that they make: there is a need to foreground the importance of scrutinizing these decisions, and not simply to consider the downstream consequences of data/digital development.

**2.3 Finding the Right Blueprint or Rethinking RCR?**

Given the potentially problematic status of data science practice, an appeal to create *ad hoc* training in data ethics is unsurprising. However, in many cases, this training was envisioned as an additional element within existing RCR programs. This positioning is logical in light of shared content and budgets. However, it is also potentially problematic in terms of content delivery. Current RCR programs may not be reliable blueprints for data science, at least for two reasons.

First, the paradoxes and tensions noted above may lead RCR programs for data scientists to present ethics as a sterile and bureaucratic ritual. In fact, using current RCR programs as models for data science ethics training may mean wasting a unique opportunity to design a different approach to RCR training. The value-ladenness and unanticipated effects of data science provide fertile ground for foregrounding the aspirational attitude that motivated the creation of RCR programs decades ago. Because the daily activities of data scientists are, to use Vallor's expression (2016), 'techno-moral', there is a critical need for RCR programs to empower data scientists to realize the aspirational expectations that motivate their research. Moreover, given the highly discretized and iterative nature of data science, linking ethical decisions to highly repeated tasks affords the opportunity to design training that supports the internalization of norms by means of habituation. This would address the often-cited (but questionably realized) aim of RCR to foster "research integrity through changing behavior, character, and judgment" (Chen 2021, p. 232). Ethical choices, whether practitioners are aware or not, are made at every single step of the research lifecycle. However, few ethics training curricula explicitly link content to these repeated daily actions and thus fail to provide the cognitive links between individual practice and the broader ethical and societal relevance of technical decisions.

In addition to the mismatch between aspirational motivation and compliance-based implementation, linking data ethics content to current RCR programs may not be ideal for data scientists for *structural reasons*. Current iterations of RCR training make use of certain aspects of the framework to demonstrate ethical behavior in daily research practices. Utilizing these "daily research examples" is critical to fostering self-reflection amongst the learners, but tend to focus on more traditional research ethics case studies such as working with human or animal subjects, or FFP misconduct. Many data scientists are removed from the collection of primary data, and as a result examples about human/animal research will have little resonance. Furthermore, as data science practitioners are highly heterogeneous in their background and roles (e.g. researchers, technical support, data curators), many will not be focused on producing academic writings as their primary outputs. This decreases the impact of FFP and responsible authorship training.

While this section notes considerable challenges for integrating data ethics into RCR training, it also draws attention to the opportunities inherent in this field of practice. Most importantly, it highlights that data scientists, regardless of their role within the research process, will be engaged in highly repetitive tasks while engaged in key strategic decisions regarding the selection of data, the design of algorithms and the development of computational models. In the next sections we describe how these characteristics can inform innovative RCR training for data science practitioners.

## 3. WHAT SHOULD A RCR PROGRAM FOR DATA SCIENTISTS LOOK LIKE?

The previous sections outline two key requirements for an effective RCR program for data scientists.

First, the programs must orient around the aspirational motivations of RCR. This means that the opportunity of addressing ethical and societal issues raised by data science tools should not be missed by superimposing old and ineffective RCR programs. What is needed is for RCR programs to be *transformational* in what students will learn. As a technical course in one of the aspects of data science might be transformative by teaching a skill and providing the bases to further cultivate that skill, so an RCR module can be transformative by providing the bases for cultivating skills or traits to identify the moral and political dimensions of micro-technical choices. The 'transformative' in RCR modules must relate to character development,[3] equipping data scientists with the instruments for effective decision-making.

---

[3] We use the term "character development" from a virtue ethics perspective to denote the ability of an individual to cultivate traits that will enable them to identify and perform the right action within a specific context. Character development also includes developing a consistency in action, so that the

The second important consideration is that any RCR program for data scientists should not create the impression that ethics is external to the practice of data science. While some of the ethical and societal challenges described in Section 2.2 are clearly internal to the practice of data science, the mode of their presentation can nonetheless fuel 'compliance' rather than critical self-reflection and awareness. Ideally, a successful RCR program teaching ethics as internal to data science practice will provide the basis for a 'data science practical wisdom' or 'data science phronesis'[4] which augments the perception of the ethical imports of data science procedures, well beyond the artificial settings provided by RCR classrooms where usually only cases that blatantly undermine research integrity are discussed. It is important to stress the fact that we aim to provide just the 'basis' of practical wisdom. One may argue (and with good reason) that the limited duration of a RCR program will not allow the cultivation of something like practical wisdom, and this is because of the nature of virtues like phronesis. Notoriously, what counts as a virtue is a matter of much controversy in virtue ethics (Vallor 2016). Sometimes scholars refer to excellences, other times to character traits, and sometimes to reliable dispositions (Annas 2011). But all these accounts share the idea that a virtue is something stable, that is cultivated through practice and by following exemplars, and that is entrenched in the character of individuals. This means that to cultivate virtues one needs time, and hence it is unlikely that in the space of one module one will cultivate a full-blown virtue. This is certainly true, but it is not a problem at all. In fact, the same challenge applies to skills, which are acquired in the same way virtues are (Russell 2015). However, this is not seen as an argument against attending technical courses which provide some technical background and lots of practical activities. While these courses are not meant to have the students to cultivate fully the technical skills, they are seen as a way to familiarize them with certain practices and to develop certain proto-virtues that we call 'preliminary abilities', which can become virtues with more practice. We think about RCR modules in the same way, namely as a strategy to familiarize students with the moral and political dimension of seemingly neutral technical activities. This is why above we have said 'the basis for practical wisdom', rather than full-blown practical wisdom.

---

individual will be able to apply the same reasoning (and enact the same virtuous behaviour) regardless of the context.

[4] The concept of "phronesis" relates to the consistency in behaviour mentioned in footnote 3 above. Phronesis, or practical wisdom, is the trait cultivated by mature virtuous individuals and refers to their ability to identify the right action within specific context. Phronesis involves not only that the individual possesses the virtues necessary for the action, but that they have a critical understanding of the context in which they are so as to identify what is the "right" or "appropriate" action in that specific setting.

These requirements delineate an ideal of RCR programs as 'transformational' and 'internal to the practice'. In order to make this ideal more concrete, we propose a RCR program with some notable characteristics.

First, the program must be embedded (Grosz et al 2019; Bezuidenhout and Ratti 2020; McLennan et al 2022) in technical curricula. There is a well-documented and positive trend of developing such 'embedded' programs, with notable examples including Harvard University[5], MIT[6], Technion, University of Toronto[7], and Stanford University[8]. Embedding ethics in technical curricula provides the opportunity to teach ethics exactly where the 'action' is taking place. By showing a more direct connection between technical choices and ethical relevance, ethics is internalized as part of the daily practice of data science. Embedding an RCR program into technical curricula requires that RCR modules must be specific to the kind of tasks that characterize the specific courses they are embedded in. One way to do this is by reflecting the highly discrete and repetitive nature of the data science pipeline, such that the modules make clear how responsible data science is implicated in operations such as formulating a data science problem, selecting data sets, preparing data sets, etc. More precisely, one can use the idealized schema formulated in (Ratti and Graves 2021), and think about the data science process as a iterative process consisting of seven stages grouped in three main phases. The first phase consists of stages involving problem definition, and data acquisition. The second phase includes the stages of data understanding and preparation, data analysis and modeling, and validation and interpretation of the model itself. Finally, the last phase includes the deployment of the model, and the evaluation of feedbacks. Ideally, one must be able to devise a RCR module for each of these stages.

Next, the embedded curricula should also be *transformational* in the sense that they must be scaled-up to all technical courses that data scientists must take during an undergraduate or graduate curriculum.

Furthermore, RCR programs for data science must take the form of ethics exercises that habituate and familiarize students with ethical reflections on responsible conduct within the context of specific tasks. This habituation would involve developing the practice of considering actions from an RCR perspective until this practice becomes a habit, and an unconscious part of planning research activities.

---

[5] https://embeddedethics.seas.harvard.edu/
[6] https://computing.mit.edu/cross-cutting/social-and-ethical-responsibilities-of-computing/
[7] https://www.cs.toronto.edu/embedded-ethics/
[8] https://ethicsinsociety.stanford.edu/tech-ethics/tech-ethics-center-initiatives

Fourth, the curriculum should equip students to deal with conducting data science research/practice in a variety of different contexts. Instead of exposing students to rules established within a specific institution, the curriculum should help them to identify the moral and political relevance of in a variety of data science contexts . Students should be able to critically understand, and cope with, the influence that the context has on their ability to enact ethical decision making and practice.

The connections between ethics, habituation, and daily tasks lead to our proposal that RCR programs for data science will be optimally transformational and internal if oriented around the concept of *microethics* (Komesaroff 1995). This concept emphasizes the importance of incremental and micro decisions and their ethical relevance. By fostering self-reflexiveness on the ethical implications of each micro-task – introduced as a component of each course of a technical curriculum – microethics makes ethical reasoning a familiar activity. This approach offers "a way of developing a multifaceted awareness of the contexts of micro-decision" (Bezuidenhout and Ratti 2020, p. 8) and cultivates a much-needed moral sensibility.

## 4. CASE STUDY: CODATA-RDA SCHOOLS FOR RESEARCH DATA SCIENCE

In order to demonstrate how our proposed curriculum of data science-focused RCR training can be operationalized, we include the case study of the CODATA-RDA Schools for Research Data Science (SRDS)[9] where both authors teach RCR. The SRDS network was founded in 2016 as a means of providing basic data science training to early career researchers (ECRs), many of whom are based in low/middle-income countries. To date, 16 schools have educated over 500 alumni through two-week residential schools held around the world. The curriculum covers software/data carpentry (R, Git, data visualization) as well as topics such as data security, neural networks, and computational infrastructures. It also includes research data management, Open Science, responsible authorship, and data science ethics (Bezuidenhout, Quick, and Shanahan 2020; Bezuidenhout et al 2021).

The data science ethics training (RCR and Open Science) receives five hours of formal instruction time during the course of the curriculum. This means that, like many other RCR courses, there is little time for long-term engagement and in-depth ethics instruction. The challenge for the instructors is thus to introduce students to key ethical concepts and current ethical debates in data science in a way that ensures they understand their own ethical responsibilities and how these responsibilities are enacted in daily research practice.

---

[9] https://www.datascienceschools.org/ (accessed 02/02/2023)

Unlike many other ethics short courses, we opted not to utilize high-level case studies as a primary means of instruction. While engaging for the students, we felt that these case studies often do not provide the cognitive pathway through which students can locate their own actions within the evolving ethical landscape. Presenting students with the 'big picture', or asking them to occupy hypothetical roles can provide the misimpression that ethical dilemmas only operate on societal levels or at higher levels of the career trajectory.

The SRDS curriculum utilizes the formal teaching time to introduce ethics, RCR and Open Science to the students, and to highlight the key responsibilities and characteristics of a responsible data scientist. A later session also introduces topical ethical issues within data science. The different themes of the lectures are then practically introduced through a series of ethics exercises linked to the technical module content. These ethics exercises are run throughout the school after each module and serve as a means of not only foregrounding individual responsibility but also iteratively repeating the concepts from the formal lectures. Together, we call this curriculum "open and responsible (data science) research" (ORR).[10]

Our idea of an embedded, transformational, incremental, and microethical RCR training is operationalized in the context of CODATA-RDA through five stages of cognitive linkage. First, instructors give students a general understanding of what ethics and moral abilities can possibly be. This general understanding should not be too technical in terms of meta-ethics or moral philosophy, as the main goal is to communicate how moral and political aspects are playing leading roles even in technical decisions. For this reason, the second step is to understand how technical decisions and ethics are connected in the data science environment. This is done by problematizing the operations taught in the technical modules into microtasks, and understanding their connections to the proto-virtues like moral abilities we refer above, and ethics: what is the relation between microtasks and virtuous behavior? The third step is to connect the microethics of technical tasks to the working environment. In this way, virtuous and vicious behavior is not only analyzed in terms of isolated microtasks; rather, the role of the environment in shaping the behavior of individuals is accounted for. In this context, vicious behaviour is understood as behaviour that either undermines the integrity of the researcher or the research. After taking stock of the complex relation between microtasks, the environment, and virtuous/vicious behavior, the next step is to discuss how microtasks and the environment can be modified - in other words, the fourth step is to develop resilience strategies. Finally, the fifth step is to discuss how to identify mentors and exemplars in one's own environment, and how to develop strategies of resilience to cope with (un)expected environmental challenges to open and responsible data science practice. In the next sections, we describe each of these five steps in detail.

---

[10] An example of the 2022 programme for the SRDS school run at the ICTP in Trieste can be found here https://indico.ictp.it/event/9806/other-view?view=ictptimetable

## 4.1 First Step: A General Understanding of Ethical Issues in Data Science and Open Science

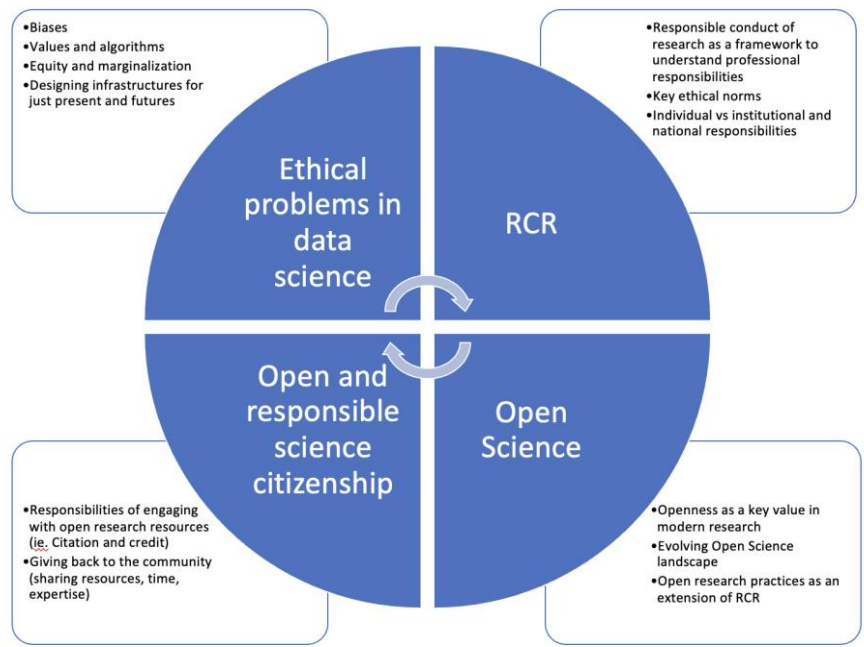

Figure 1: Overview of topics taught in formal ethics lectures at SRDS

The first step in ORR training is to introduce students to ethical concepts and issues surrounding data science, RCR, and the culture of open science. The topics covered in these formal lectures are illustrated in Figure 1 above. The first two topics provide an introduction to RCR and Open Science. These concepts are linked by illustrating how open research practices address key activities identified in RCR, which is an important means of ensuring that students do not compartmentalize the different movements, meaning that they should not think of them as unconnected or independent of each other. Both the RCR and Open Science topics highlight not only the responsibilities of the individual researcher, but also the roles that institutions and national and international infrastructures play in safeguarding research integrity.

The third topic brings the responsibilities identified in RCR and Open Science together under the concept of 'open and responsible science citizenship'. The concept of science citizenship has proven a useful means of ensuring that students are able to personally contextualize the earlier topics. Most of the students taught on the SRDS have no prior ethics training and little knowledge of Open Science. Because of this, these topics - together with the extensive lists of expectations outlined - can seem overwhelming. Bringing the discussion on responsibilities back to citizenship is useful, as all learners have some understanding of the concept. We use this as a means to discuss community membership, community rules, and the rights, freedoms and duties associated with citizenship. This is a means of discussing various duties, such as community rules (i.e., crediting and citation), duties (i.e., peer review and contributing to community activities) and community maintenance (safeguarding open resources and whistleblowing).

The last section taught in the formal lectures applies the concept of science citizenship more broadly to look at science in society. In this section we cover topics such as bias and value-ladenness in algorithms, data visualization and the problems of marginalization and misrepresentation through data re-use, data infrastructures and information and communication technologies. This section serves to orient the students towards the societal impact of data science but also to highlight their responsibility for engaging (as data experts) in these discussions.

**4.2 Second Step: Linking Ethical Content to Data Science Tools**

The second step is to understand the connections between ethics and technical decisions. In the SRDS we run a series of ethics exercises through which students can connect daily research decisions, and the technical tools used in these daily practices, with RCR ethics, as well as the "big picture" ethics discussions. This is an important means through which to iteratively revisit ethics throughout the summer school, but also to ensure that students do not compartmentalize ethics as a "stand alone" topic that is unrelated to the technical subjects being taught at the school.

The ethics exercises that are used to understand the connections between ethics and technical decisions consist of problematizing the practice of data science into discrete microtasks that make use of data science tools. The ORR microethics exercises are designed to accompany each technical module of the SRDS curriculum, and foster reflection on the use of the data science tool and the broader implications of misapplication. A full list of ethics exercises utilized at SRDS is published online[11], with an example given below.

## GitHub: Limits of Openness (A)

Due to U.S. trade controls law restrictions, your GitHub account has been restricted. For individual accounts, you may have limited access to free GitHub public repository services for personal communications only. Please read about GitHub and Trade Controls for more information. If you believe your account has been flagged in error, please file an appeal.

- GitHub is blocked to researchers living in countries under sanction from the US. This means that they are not (legally) able to collaborate on this platform or to benefit from the vast resources it offers. Is this a reason for you not to use this tool?

Figure 2: example microethics exercise from GitHub module

GitHub is a platform for (amongst other things) sharing software and managing version control. Data science students will interact with GitHub on a regular basis as they share code, reuse code and engage with scholarship in their field. Nonetheless, the use of GitHub is not unproblematic. As demonstrated in figure 2 above, GitHub is blocked in countries against which the US holds sanctions. GitHub thus provides a useful way of linking issues of inclusion and exclusion to daily research practice, and the future use of the platform provides a means through which students can regularly question whether they are fulfilling their duties towards the inclusivity and universality of research. When students use the platform, they are encouraged to think about whether there are alternatives actions that they can undertake to minimize the harm caused by the geoblocking of GitHub.

---

[11] 10.5281/zenodo.6874958

Students undertaking the GitHub ethics exercise illustrated in Figure 2 above are asked to vote, choosing between the following options:

- no, it's not my problem
- no, the blocking is probably justified
- yes, but only if I am collaborating with researchers in those countries
- yes, it is unjust to shut researchers out of Open Science platforms

The results of the voting provide a basis on which to start a discussion about key issues, such as limits of openness, equity and marginalization, political influence on research and non-maleficence in research. This provides the students with a cognitive link between the ethics content from the lectures and its applicability to the data science tools that are used.

In addition to providing the cognitive links between daily research practice and broader ethical discussions, the microethics exercises offer an important means of distinguishing between good and poor conduct. In the discussions it is also possible to critically unpack what 'misbehavior' would look like within a specific microtask. The identification of these poor conduct enables the students to identify what the right action would be. Table 2 below illustrates how poor conduct can be identified from the specific microtask under discussion within the ethics exercise.

As will be apparent to many, these discussions are underpinned by the virtue ethics traditions of virtuous and vicious behavior. Misbehaviors lie at two extremes - omission and commission - Understanding what poor conduct would look like within a specific context, and consequently the 'good behavior' requires finding the midpoint of good behavior between these poles. Using these vices of omission and commission as prompts within discussion assists the lecturer in broadening out the students' understanding of poor research conduct.

Highlighting the distinction between doing too little or doing too much (as demonstrated by table 2) is an important contribution to the ethics exercise discussions. Indeed, it contrasts to the highly bureaucratic approach of other RCR trainings, which provide a list of rules to follow, but do not encourage discussion around different ways of enacting poor conduct around a single action.

| If I'm prevented from doing my task well, what happens? | Microtask | If I am under pressure to finish the task quickly, what happens? |
|---|---|---|
| If I don't have training or | I use R to visualise data | I may not be scrupulous in |

| | | | | | |
|---|---|---|---|---|---|
| software I may construct misleading visualizations | | | | checking my visualization or may not seek the advice that I need |

Table 2: sample of instructor sheet with prompts for guiding behaviour outlining potential vicious behavior associated with a microtask. By vicious behaviour we refer to poor conduct within the situation.

This strategy will be familiar to all scholars using virtue ethics or virtue language. Indeed, following the virtue ethics tradition, the discussion aims to provide students with the tools to identify the *right action* to take in relation to the use of tools within their own research context. In effect, for students to get used to identifying the "golden mean" of good behavior when they utilize data science tools or engage in other research practices.

Table 3 provides even more context underpinning these discussions, and is used as a train-the-trainer tool for teachers undertaking to teach Open and Responsible Research at the SRDS. As is evident, the virtues associated with the "right" execution of the microtask are counterbalanced by the vices of omission and commission that are linked to the misbehaviors identified in table 2 above.

| If I'm prevented from doing my task well, what happens? | Vices (omission) | Microtask | Virtues | If I am under pressure to finish the task quickly, what happens? | Vices (commission) |
|---|---|---|---|---|---|
| If I don't have training or software I may construct misleading visualizations | Slothfulness, Apathy | I use R to visualize data | Temperance, wisdom | I may not be scrupulous in checking my visualization or may not seek the advice that I need | Recklessness, Lack of judgment |

Table 3: extended mapping of virtue ethics guiding discussions on microethics

By following the structure of the ethics exercise discussions, the instructor is thus able to introduce the students not only to virtues, but also to the process of identifying the "golden mean" between vices of omission and commission that is suitable for a specific context and action. It is important to mention, however, that the formal language of virtue ethics is rarely used within the SRDS school. As most of the students come from natural and life science backgrounds, the introduction of a list of new terms can cause cognitive dissonance and detract from the learning process. Rather, virtuous behavior and associated vices are termed using the conventional "right behavior" and "misbehavior", providing consistency between our teaching and RCR instruction more generally.

It is also important to spend a few words on the process of identifying the 'golden mean'. This is connected to a central trait in our RCR training: the virtue of phronesis. Phronesis has many names, including prudence or practical wisdom. It is a virtue or habit that is not just limited to the Western tradition starting from Aristotle, but it is also recognized in the Confucian moral tradition as well as in Buddhism (for a detailed comparison of Aristotelian phronesis with these traditions see Vallor 2016). Phronesis is defined as the necessary virtue to identify the right action within a specific context, in the sense that 'controls' the "enactment of a person's individual moral virtues, adjusting their habitual expression to the unique moral demands of each situation" (Vallor 2016, p. 19). Phronesis is deliberative in the sense that is operationalized via prudential judgements. But despite the importance of a skill such as 'prudential judgment', phronesis is not only the ability to "judge and choose a course of action" (Vallor 2016, p. 106); it is much more, as it also presupposes other moral virtues, but at the same time the other virtues cannot be properly coordinated and modulated without phronesis. When describing 'technomoral' wisdom, Vallor notices that (practical) wisdom is not a specific excellence like other virtues, but rather is a general condition of possibility for the other virtues to be exercised - it is unifying other virtues. The goal of the ethical exercises is to put students on the right path of a moral self-cultivation that can possibly lead to practical wisdom; this 'wisdom' will allow them to aim at the correct goods (i.e., the ones which lead to human flourishing within a certain context) and for the right motives. While we are certainly aware that a school cannot result in cultivating such an important and controversial virtue, we do believe we should start from somewhere, and SRDS can be a much needed starting point.

### 4.3. Third Step: Recognizing Context

In order to assist the students in understanding the role of the research context on their conduct, the ORR curriculum includes a session dedicated to discussing the challenges to responsible research conduct that students identify in their home environments. The object of this session is twofold. First, to demonstrate to the students that they are not alone in having problems and concerns about implementing open and responsible research practices, and second that many of these problems can be addressed using national and international infrastructures, tools and communities.

In structuring these discussions, we draw on the Institutional Analysis and Development Framework developed by Elinor Ostrom (2007). This framework analyzes institutional structures based on biophysical conditions, attributes of the community and rules-in-use. This enables institutional outcomes to be understood on a multidimensional level, and draws attention to the manner in which these different elements shape action situations and outcomes.

| Context category | Sub-category | Description | Characteristics of own institution |
|---|---|---|---|
| **Biophysical** | | **Access to resources** | |
| | Common resources | What resources do you have access to in order to do your work? Is anything missing? | |
| | Public resources | How is the provision of the internet, power, working environment in your institution? | |
| | Personal goods | Do you have access to a mentor, support system at work? | |

Table 4: sample of instructor sheet with prompts to introduce the exercise on identifying contextual challenges to open and responsible research practice

This exercise is introduced by the instructor, who explains the different aspects of the institution that can present challenges to open and responsible research practices (using prompts such as those outlined in table 4). The students are then invited to anonymously contribute to a shared document in which they list their concerns according to three categories:

- institutional/cultural issues
- infrastructural issues
- personal concerns

Discussion of these concerns forms the basis of a broader discussion about challenges and existing opportunities for their amelioration. As mentioned above, a key objective of the discussion is to empower the students with the realization that they are not alone in their concerns, and that there are structures existing that can offer support. In order to reiterate that all researchers have concerns, students are also introduced to the findings of a number of international surveys, such as the State of Open Data run annually by FigShare.

Examples of the content discussed in these exercises is available online[12], while a sample of the concerns and possible solutions are presented in table 5 below.

|  | **Concerns** | **Possible Resources to Use** |
|---|---|---|
| Institutional/cultural concerns | Wide mentoring gap between professors and young researchers | Look for mentors outside of institutions. This could include independent networks such as AuthorAid, OpenLifeScience, disciplinary communities or international organizations such as the Research Data Alliance. |
| Infrastructural concerns | Lack of infrastructures and institutional support on Data Archiving | Understanding international repository landscape can assist (ie. re3data), as well as understanding what constitutes a *trusted digital repository* and open licensing for research resources. Many institutions are developing institutional |

---

[12] https://zenodo.org/record/6875154#.Yt-qOC98rBI

| | | repositories. Enquire from your library services about current developments. |
|---|---|---|
| Personal concerns | Limited knowledge of open science infrastructures | Considerable Open Science online training such as Open Science training handbook, Open Science MOOC and FOSTER. Other useful courses through data.europa academy. |

Table 5: sample of concerns (column 2) raised by students in 2022 SRDS schools and group discussion on potential tools to ameliorate concerns (column 3).

As will be evident, both from table 5 and the extended document, is that we strongly encourage our students to seek advice, develop strategies of support and identify mentors. The identification of networks of support and trusted advisors is key to developing the ability to identify the right action within a specific context. Crucially, the exercise highlights the importance of looking for support beyond their researchers' institutions. There are a range of local/national/international communities that provide these levels of interpersonal contact, and having the students understand these opportunities is vital for their ability to withstand the environmental challenges to responsible and open research practices that will inevitably occur during their careers.

## 5. STRATEGIES OF RESILIENCE

The different steps of the ORR curriculum are outlined in Figure 2 below. In contrast to other RCR curricula, our training focuses on linking ethical principles and current broader ethical concerns in data science to the daily research practices of the individual researcher. This empowers students to realize that their actions cumulatively play a role in safeguarding beneficial research and just digital futures. The curriculum also focuses on problematizing the research contexts in which the students are embedded, focusing on helping them to find the *right action within the context*. This marks a change from RCR education that utilizes rules or codes of conduct as a primary means of outlining good conduct. The focus on right action in context reflects our awareness that RCR readiness differs greatly between institutions and that there cannot be a "one size fits all" approach to outlining what is the best course of action within a research context. Rather, the ORR curriculum focuses on getting students to be mindful of their responsibilities and to critically find tools to help them implement them within their research context, while also offsetting the impact of negative environmental influences.

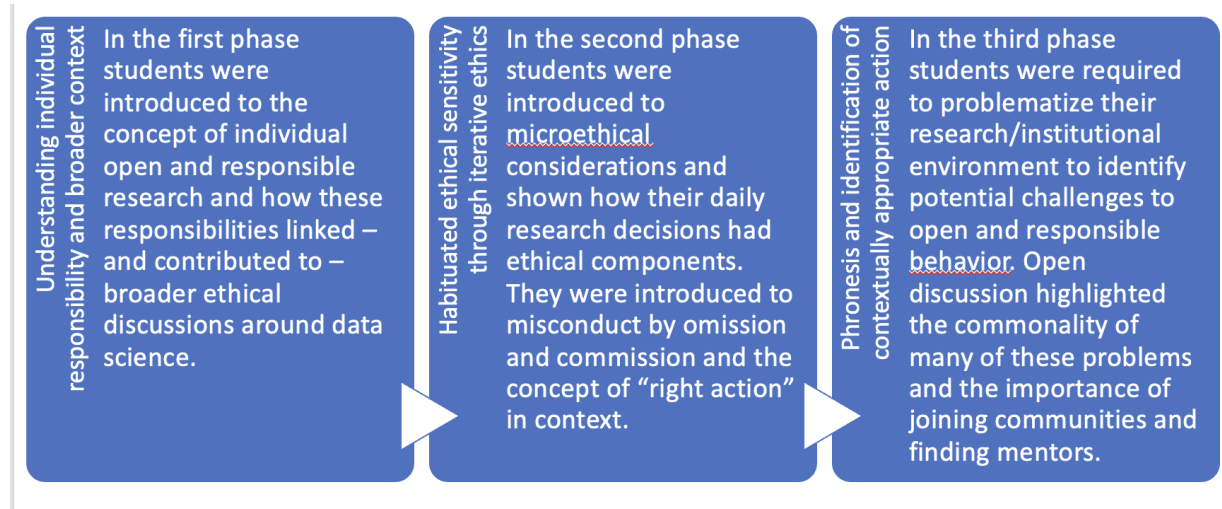

Figure 2: Building up resilience through ethics training

The discussions about research contexts, together with the discussions associated with the microethics exercises also serve another important purpose. By encouraging students to discuss the challenges they face in their research environments, we create a space in which students recognize that their concerns are in no way unique. Students around the world struggle with similar issues relating to implementing responsible and open research practices. In this way, we draw the students' attention to the importance of integrating into local/national/international research communities and other organizations offering support. This is reflected in the sample answer outlined in table 5.

Taken together, the ORR curriculum offers students a pathway towards developing not only open and responsible research practices, but also strategies of resilience in the face of (un)expected challenges. This pathway is outlined in Figure 3 below. Providing students with these strategies of resilience is an important contribution towards long-term ethical behavior, and ensures that students do not become discouraged as they go forward as early career researchers in environments that are perhaps not optimally equipped to support an "ideal" of RCR practice.

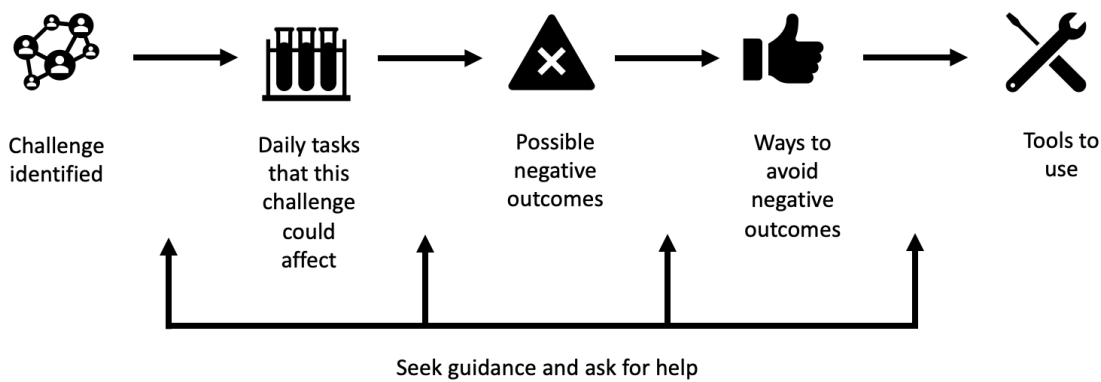

Figure 3: Resilience strategy logic

Developing the ability to identify "right action", together with developing strategies of resilience to deal with (un)expected challenges are an important counter to the rule-focused approach of many RCR training. If students are simply taught to follow rules - and that this exemplifies good conduct - they are ill-equipped to deal with situations in which these rules cannot be followed in the ways in which they were taught. As few students will remain in their original institution for the duration of their career and often move between countries or sectors, this should be reason for concern.

# 6. CONCLUDING COMMENTS: LEVERAGING DATA SCIENCE TO FOSTER ETHICALLY ROBUST DIGITAL SYSTEMS

The expansion of ethics training, such as ORR, that focuses on contextual awareness and strategies of resilience has the potential to positively impact data systems and digital infrastructures. Fostering data scientists who are habituated to scrutinize the ethical impact of the smaller units of digital infrastructures and the practices of data use will ensure that the developments within the emerging digital/data landscape are continually monitored.

We outline this aspiration in figure 4 below. We suggest that the training of data scientists in ethical awareness and resilience will support them in becoming virtuous and able to identify right action in context. This will lead to the development of ethically robust data systems that, in turn, influence users.

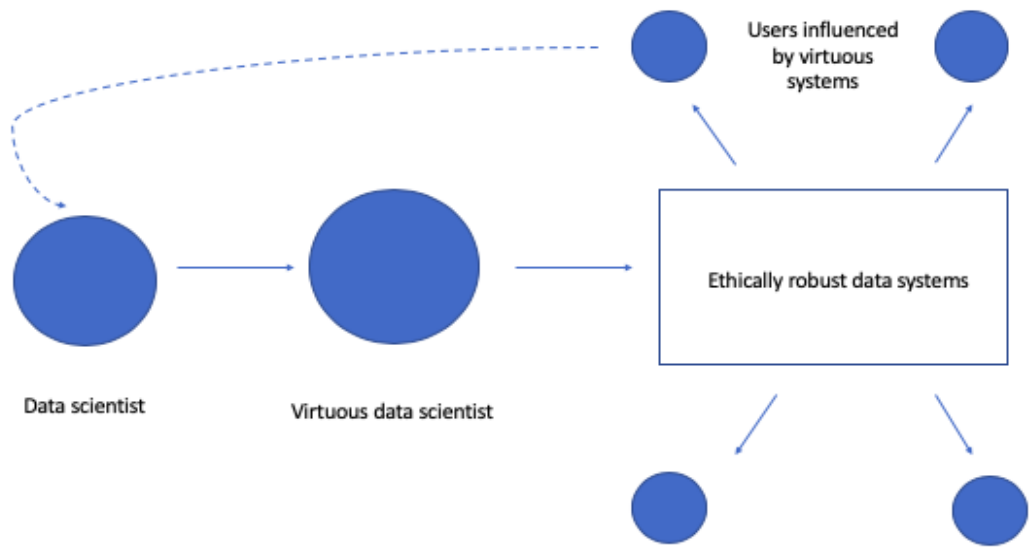

Figure 4: Developing ethically robust data systems

Together, it is possible that we will globally be able to ensure that our digital futures reflect the ethical principles that underpin research.